\documentstyle[12pt,aasms4]{article}

\slugcomment{submitted to The Astronomical Journal}

\lefthead{Richer et al.}
\righthead{Messier 4 Response}

\begin{document}

\title{Concerning the White Dwarf Cooling Age of M4:  A
Response to the Paper by De Marchi et al. on "A Different Interpretation of Recent HST Observations"$^1$}

\author{H.B. Richer, J. Brewer, G.G. Fahlman$^2$, J. Kalirai}
\affil{Dept. of Physics and Astronomy, University of British Columbia}
\centerline{richer/jbrewer/jkalirai@astro.ubc.ca, greg.fahlman@nrc-cnrc.gc.ca}

\author{P.B. Stetson}
\affil{National Research Council, Herzberg Institute of Astrophysics
         }
\centerline{peter.stetson@nrc-cnrc.gc.ca}

\author{B.M.S. Hansen, R.M. Rich}
\affil{Department of Physics and Astronomy,
           University of California at Los Angeles
           }
\centerline{hansen/rmr@astro.ucla.edu}

\author{R.A. Ibata}
\affil{Observatoire de Strasbourg}
\centerline{ibata@newb6.u-strasbg.fr}

\author{B. K. Gibson}
\affil{Centre for Astrophysics and Supercomputing, Swinburne University}
\centerline{bgibson@astro.swin.edu.au}

\author{M. Shara}
\affil{American Museum of Natural History
           }
\centerline{mshara@amnh.org}

\altaffiltext{1}
{Based on observations with the NASA/ESA Hubble Space Telescope, obtained at
the Space Telescope Science Institute, which is operated by AURA under NASA
contract NAS 5-26555. These observations are associated with proposals GO-5461 and GO-8679.}
\altaffiltext{2}
{Herzberg Institute of Astrophysics}

\begin{abstract}
We respond to the
paper by De Marchi, Paresce, Straniero and Moroni (2003) on the white dwarf cooling age of M4.
The authors question the data analysis and interpretation that
led to the conclusions in Hansen et al. (2002).
In their paper, De
Marchi et al. are unable to obtain photometry as deep as ours from the
same data set and therefore assert that only a lower limit to the white dwarf
cooling age for this cluster of $\sim 9$ Gyr can be obtained.

In this short contribution we show that shortcomings in the data
analysis and reduction techniques of De Marchi et al. are responsible for their
inability to reach the photometry limits that our study reports. In a forthcoming paper 
in which the complete techniques for age determination with
white dwarfs are laid out,
we demonstrate that their method of fitting the luminosity
function gives a spuriously low white dwarf cooling age.

\end{abstract}


\section{Introduction}
In early 2001, the Hubble Space Telescope obtained the deepest data
set ever observed for a globular cluster, an exposure of 123 orbits
centered on a field 5$\arcmin$ from the center of the globular cluster
M4.  We reported analysis of the lower main sequence of M4
(Richer et al. 2002) and a determination of the white dwarf
cooling age of M4 (Hansen et al. 2002).  The latter is derived
from a $\chi^2$ fit of models to the entire white dwarf luminosity
function. The derived age of
12.7$\pm 0.35$ ($1\sigma$ statistical error only) is consistent with other age determinations for
old star clusters, and with the {\sl WMAP} concordance cosmology
age (Spergel et al. 2003).

In a recent contribution, De Marchi, Paresce, Straniero and Moroni (2003) performed
 a reanalysis of
the long exposure data from our HST project GO-8679 (Richer et al. 2002, Hansen et al. 2002) 
and were unable to reproduce results
similar to ours.
In particular their photometry failed to reach as faint as ours by almost a full magnitude. 
This can be seen quite dramatically in Figure 1 where we plot
our CMD in the white dwarf region from this data set and that of De Marchi et al. 
taken from their paper. Continuing, the authors claim that the
only possible age constraint for M4 is that the lower limit for the
white dwarf cooling age is 9 Gyr, whereas we (Hansen et al. 2002) derive
a cooling age of 12.7 Gyr.  The analysis of de Marchi
et al. raises distressing questions about our competence
both in data analysis and in interpretation.  As they have
already been cited by others in the literature (e.g., Moehler et al. 2003)
we believe that it becomes crucial for us to respond.

The conflict raises two questions.  First, why do two independent teams
approaching the same data set get such different results?  Second,
what is the correct approach to fitting a measured white dwarf
luminosity function?

We answer the first question in this contribution and the second in detail in
Hansen et al. (2004, in preparation) and illustrate how the De Marchi et al. analysis fails on both counts.

\section{Data Reduction and Analysis}

The reason De Marchi et al. are unable to reproduce the depth and quality of 
our CMD is purely one of data treatment. The HST data set under consideration, GO-8679, contains
 98 F606W and 148 F814W images each of exposure 1300 sec. A first-epoch set of images
taken in F555W and F814W (GO-5461) were obtained by us in 1995
(Richer et al. 1995, 1997) and were not nearly as deep. Some care is required in
getting the most out of this large and impressive data set.

In our case, after accounting for dithers, those frames corresponding to a
particular epoch and filter were averaged together with pixel rejection---the 
$n$ highest pixels being rejected to eliminate cosmic ray contamination.
From the HST WFPC2 Manual the mean number
of pixels on a given chip affected by a cosmic ray hit in an 1800 sec exposure
is about 20,000. Hence in a stack of $N\times1300$ sec exposures we expect $0.02257N$
pixels at a given position to have suffered a hit. Taking the dispersion
to be $\sqrt{0.02257N}$ and using $5\sigma$ rejection we find that in
combining the 98 F606W images the highest 7 pixels should be rejected while
for the 148 F814W images we should eliminate the highest 9. Our procedure
thus used 93\% of the available pixels in constructing the mean.
 
By contrast, De Marchi et al. used
image-association stacks produced by the pipeline software developed
by the Canadian Astronomy Data Centre and the ESO Space-Telescope
Coordinating Facility (Micol \& Durand 2002).  This pipeline software
essentially takes the median of the available pixel values, and is therefore
less efficient in a statistical sense than the scheme we employed
as described above.
According to 
Zwillinger \& Kokoska (2000) 
the efficiency of the median ranges from 1.000 for $m=2$, 0.743 for $m=3$, ...
0.681 for $m=20$, to 0.637 for $m =\infty $. Hence De Marchi et al.
effectively used less than 2/3 of the information available in the images.

We requested from the Canadian Astronomy Data Centre the
image association stacks that were used by De Marchi et al. and compared them to the
combined images resulting from our
analysis.  After allowing for the change in image
scale (we expanded our images by a factor of three in each dimension), the
main difference is that
images of faint stellar objects appear perceptibly more diffuse in the
pipeline generated associations. This has the effect of making the faintest stars
 difficult to detect in the images De Marchi et al. used, exactly what is seen 
in Figure 1. We illustrate this by employing
the same 2 stars De Marchi et al. used to claim that objects fainter than 
F814W = 27.0 could not be measured on our first-epoch frames. Recall that
all we seek from the first-epoch images is a position so that proper motions
can be measured. All the photometry is carried out on the longer-exposure
frames. For this reason most of our positional matching comes from the
F606W $-$ F555W pair as the first-epoch exposures in F555W are actually quite
long---31.5 ksec. Figure 2 displays these two images as we produced them with the stars discussed
by De Marchi et al. circled. Clearly the lower of the two is easily visible in both frames
 while the upper of the two is not seen in the
shorter exposure first epoch. There is, however, a simple reason for this - it is in fact
not a star at all, as our deep images clearly show it to be extended. This faint
galaxy
is thus lost in the noisier first-epoch image.
Some specific numbers for these two objects are that each is detected at a $S/N$ of about 8 on the
 long exposure F606W image, the upper one is not detected on the
shorter F555W  frame while the lower is detected at $S/N$ of about 5. The lower star
has a proper motion completely consistent with cluster membership and its
measured magnitudes are F606W = 28.48, F814W = 27.25. It is a cluster
white dwarf and as can be seen in Figure 1 we are able to measure stars
almost 0.5 magnitudes fainter than this one.

One crucial point is that our experiment matches
long exposures in F606W (127.4 ksec) and F814W (192.4 ksec) with much shorter
exposures in F555W (31.5 ksec) and F814W (7.2 ksec). The question thus arises
as to how we could measure positions in the earlier epoch data for the faintest stars 
where the $S/N$ is obviously poor. For a star at F606W = 29.0 the expected
$S/N$ (from the HST ETC) in our deep F606W frames is 6.7 while a star at this
magnitude on the shorter exposure F555W frames has $S/N$ = 2.5. This is just about
what we measured on our frames. The reason we can successfully
measure positions
of such faint objects in the first epoch are twofold.

First, we applied the
finding list from the deep frames to the shallower frames. If the background noise in an image is Gaussian, then a
$2.5\sigma$ positive deviation will occur in about 6 pixels out of every 1000.
If, in a $750 \times 750$ image (neglecting the vignetted areas at the low-x and low-y
sides of the WFPC2 CCDs), we were to mark all of the $2.5\sigma$ peaks as
detected astronomical objects, we would expect more than 3,000 false
detections.  However, if we consider only the area within 0.5 pixels radius
of an object confidently detected in the long second-epoch exposure, we
expect a probability $\sim \pi \times 0.5^2 \times {6\over1000} \approx
0.005$ of finding a positive 2.5$\sigma$ deviation which is purely the result
of random noise in the first-epoch image.  That is to say, we expect of
order 5 false cross-identifications for every 1000 correct re-detections.
Even at $1.8\sigma$, presumed re-identifications will be correct 19 times out of 20.
For this reason, the knowledge that an actual astronomical object is present
somewhere nearby based upon the long-exposure second-epoch images allows us
to be confident that most of the claimed re-detections on the short-exposure
first epoch images correspond to true re-detections.  The first-epoch
astrometric positions, then, while poorer than those of the second epoch,
are nevertheless good enough to distinguish stars that are moving with the
cluster from stars and galaxies that are not.  By contrast, 
De Marchi et al. made no use of this technique and hence the shorter first-epoch frames effectively set their limiting magnitude. We checked the above by rerunning the photometry programs on the WF3 chip using a finding list with star positions generated randomly. This list contained the same number of stars as the original
finding list.
We then selected the objects whose photometry had converged on 
all frames. The resulting
proper motion displacement diagram and CMD is shown in Figure 3 and compared with the one when the true finding list is used in Figure 4. There are numerous spurious faint matches in Figure 3, but almost all disappear when the match radius of 0.5 pixels is invoked. Those objects that lie outside 0.5 pixels are
likely random noise spikes which have no correlated positions between frames.
Figures 3 and 4 taken together show that false detections of noise spikes are not a serious source of contamination in our CMDs and that the suggestion to the
contrary by De Marchi et al. is incorrect.

Secondly, the proper motions of the
M4 stars are
actually quite large (about 1 HST pixel with respect to an extragalactic background over the 
6 year time baseline (Kalirai et al. 2003)).
While the lower $S/N$ image would not produce good enough photometry,
the $S/N$ is sufficient to give the centroid, which is crucial for
astrometry.
In Figure 5 we illustrate the quality of the proper motion
separation between cluster and field from the long and short images.
Some of the brightest stars do not
exhibit clean separation due to their near saturation. However, what is clear from this diagram is that most of the
field objects can be easily eliminated down to very faint magnitudes (F606W = 29.0) by making 
the generous proper motion cut within a total value of 0.5 pixel ($\sim 8$ mas/yr) of that of 
the mean cluster motion over the 6 year baseline
of the observations. A disturbing aspect of the De Marchi et al. paper is
that nowhere do they illustrate their displacement measurements.

The analysis techniques of De Marchi et al. compounded the effect of their somewhat lower quality images. Their approach uses aperture
photometry to measure the stellar magnitudes.  We employ point spread
function fitting, which
clearly has the advantage of just using
the pixels with the largest signal---thus reducing the sky contribution---a
critical issue when attempting to measure faint stars. The effect of this is
obvious even along the brighter portion of the white dwarf cooling sequence where the 
De Marchi et al. diagram exhibits significantly larger scatter (see Figure 1). Further, as we
mentioned above, we use the positional information from the deepest frames
 in carrying out the
photometry on the shorter exposure frames. By contrast De Marchi et al. made
no use of this information. {\it This has the effect of allowing the depth of the
first-epoch images alone to set their limiting magnitude.}

\section{Other Issues}

In a paper currently in press (Hansen et al. 2004)
we consider in some detail the host of systematic effects in the white dwarf
cooling age determination. In that contribution we will comment in detail
on some of the points raised by De Marchi et al. such as
distance modulus, reddening and mass function. Here we have only considered reduction and analysis of the data. With regards to this data we offer 
our stacked images and finding lists to anyone wanting to use them and in
Richer et al. 2004 we have published the complete photometry lists. 

 On one point we agree with De Marchi et al. -- the pursuit of globular cluster white dwarfs is an important and
interesting scientific goal. However, it is also an extremely expensive project in terms
of telescope time, and we have a responsibility to take the time to give
the analysis significant thought and effort to come up with the best 
possible analysis. We would be delighted if we could detect the truncation
of the white dwarf luminosity function, and hope to do so in future work,
but until such time we reiterate the
point that there is abundant age information in the white dwarf luminosity function
and it can be modelled.

\begin{acknowledgements}
The authors would like to thank I. King and J. Anderson for reading an early version of this paper and making useful comments. The research of HBR is supported in part by the Natural Sciences
and Engineering Research Council of Canada. HBR extends his appreciation to the Killam Foundation and
the Canada Council for the award of a Canada Council Killam Fellowship. RMR and MS acknowledge support from proposal GO-8679 and BH from a Hubble Fellowship
HF-01120.01 both of which were provided by NASA through a grant from the Space Telescope Science Institute which is operated by AURA under NASA contract NAS5-26555. BKG acknowledges the support of the Australian Research Council through its Large Research Grant Program A00105171.
\end{acknowledgements}

\clearpage

\clearpage

\figcaption{Color magnitude diagrams in the white dwarf region from Richer et al. 2002 and Hansen et al. 2002 (left panel) compared with that of De Marchi et al. 2003 (right panel) taken from their paper. In both diagrams only stars detected in images from both epochs were included so that proper motion selection could be made. No main sequence
stars are included in this version of the De Marchi et al. CMD. The Richer et al. CMD is fully 1 magnitude deeper and, significantly, exhibits much smaller scatter even at
relatively bright magnitudes.}\label{fig1}

\figcaption{A small section of the WFPC wf4 chip (17$\arcsec$ $\times$ 17$\arcsec$) exhibiting our stacked F606W (second epoch) and F555W (first-epoch) images. The 2 circled objects are two of the objects discussed in De Marchi et al. (the upper is their star 2, the lower star 3) from which they claimed that objects fainter than F814W = 27 could
not be measured on the first-epoch image. The upper object is not seen on our first epoch as it is a galaxy whose soft image is lost in the noise. The lower
of the 2 objects is a cluster white dwarf, easily seen in the first epoch ($S/N$ = 5) with measured magnitudes of F606W = 28.48 and F814W = 27.25.}\label{fig2}

\figcaption{Proper motion displacement diagram and resulting CMD from wf3 obtained when the input coordinate list for the first-epoch frames are randomly generated. The upper section illustrates the proper motion displacements in HST pixels over the 6 year baseline of the data (centered on M4)
while the lower section is the CMD. The left hand panel contains the CMD for all the stars matched within
4 pixels in $x$ and $y$ while
the right hand one is for those stars matching within 0.5 pixels. This latter criterion is what we
have used to produce the cluster CMD from which the white dwarf cooling age of M4 is derived.}\label{fig3}

\figcaption{As in Figure 3 except here the correct coordinate system is used
and the matching radius is set to 0.5 pixels. Figures 3 and 4 together illustrate that spurious detections of noise spikes masquerading as faint stars are negligible in our CMDs.}\label{fig4}

\figcaption{Total proper motion difference between M4 and background/foreground objects as a function of magnitude for all the stars in the M4 images. In this diagram the motion is zero-pointed on M4 and the displacement is given in HST pixels over the 6 year baseline of the observations. The M4 stars are those
along the bottom while the field objects scatter above. The M4 internal motion is not resolved but the dispersion in the field is. In order to separate the cluster from the field all objects possessing motion within 0.5 pixel of that of M4 were assigned cluster membership. It is clear from this diagram that proper motion selection can be made down to a magnitude as faint as F606W = 29.} \label{fig5}

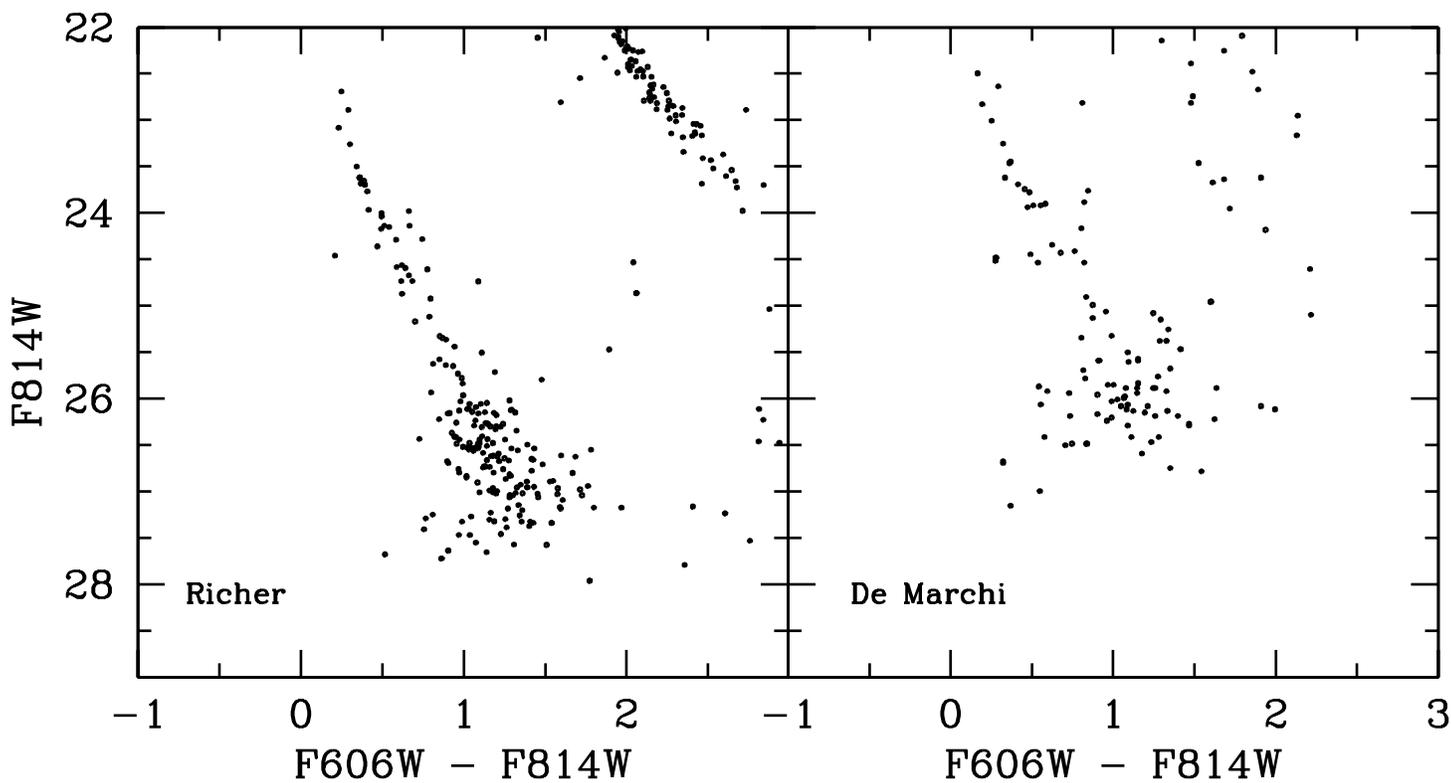
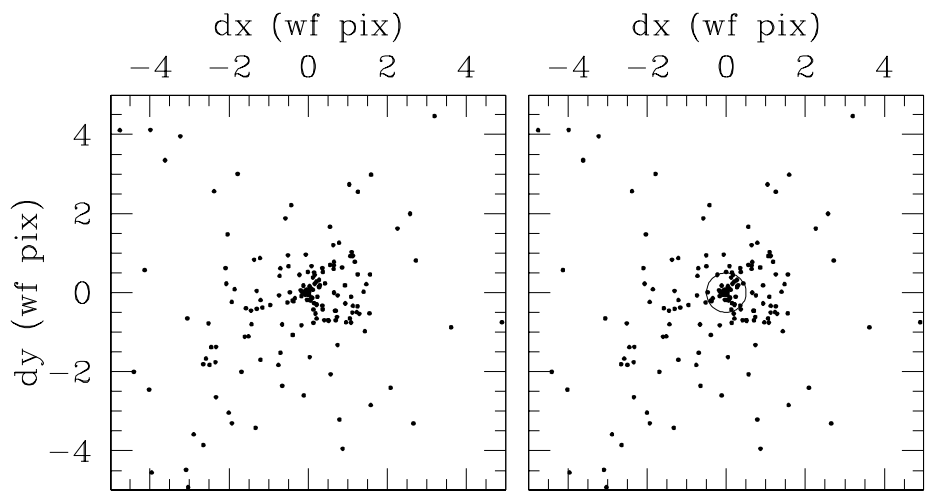
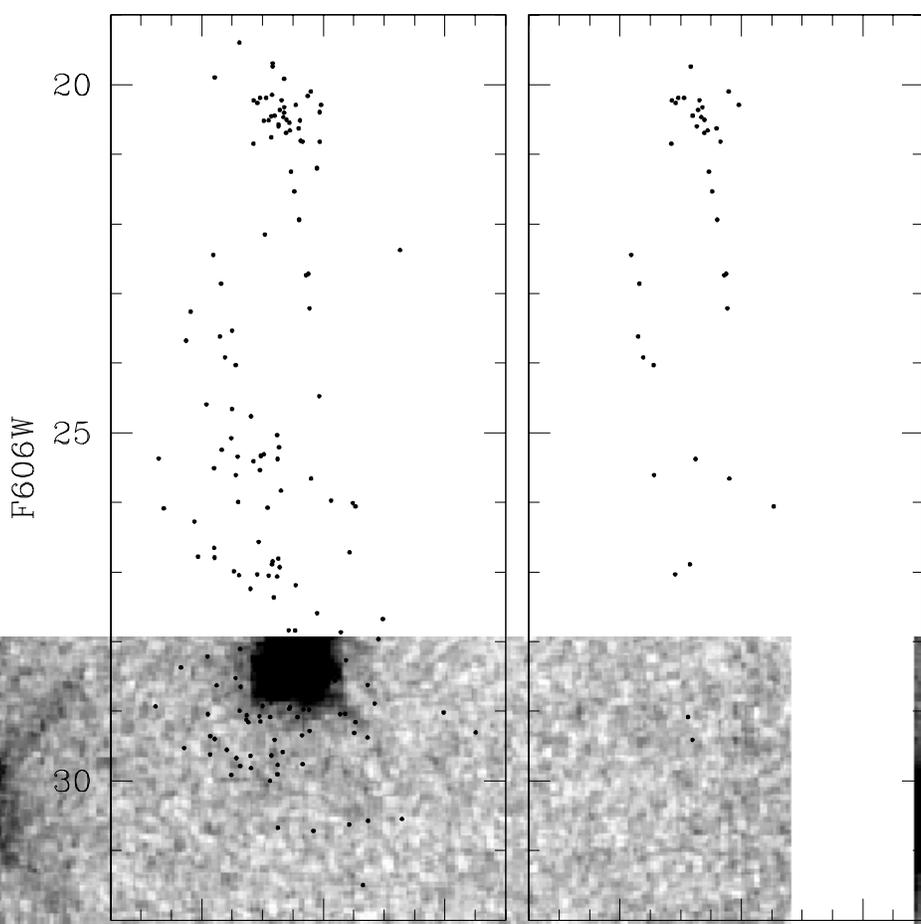
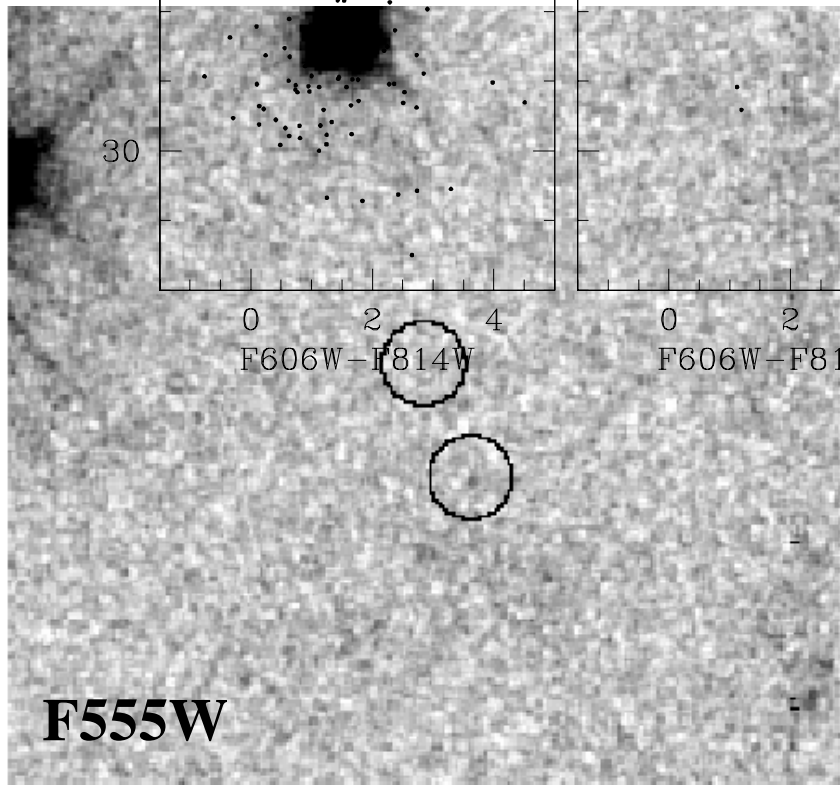
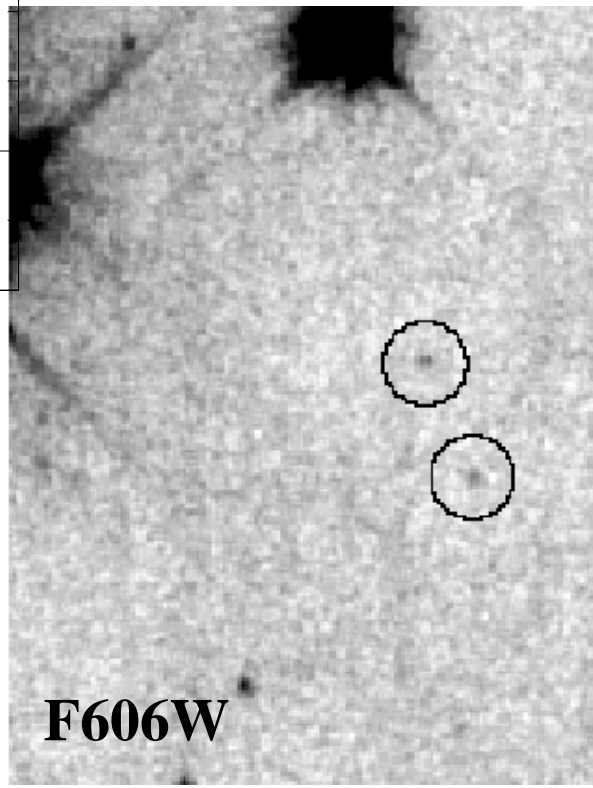
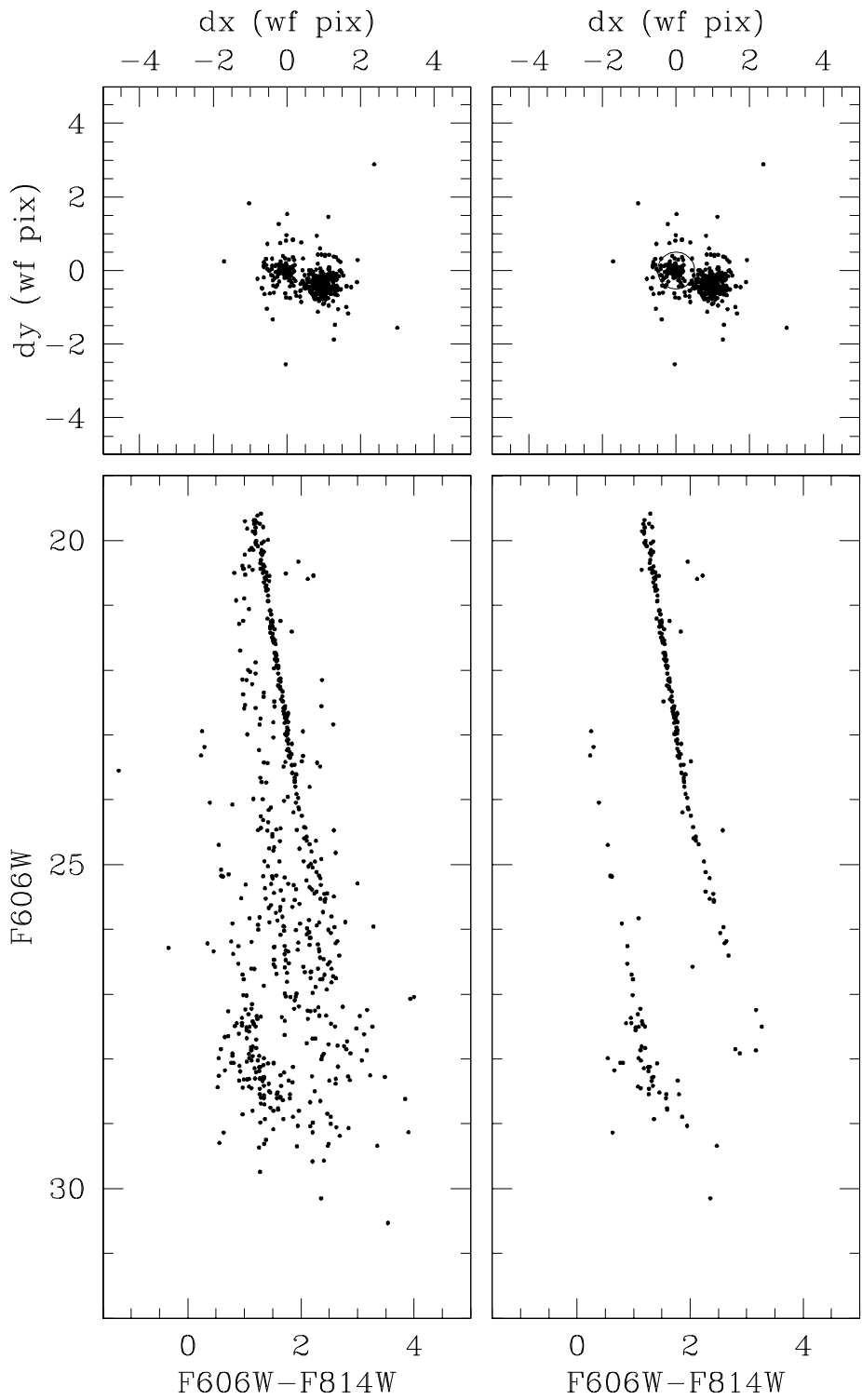
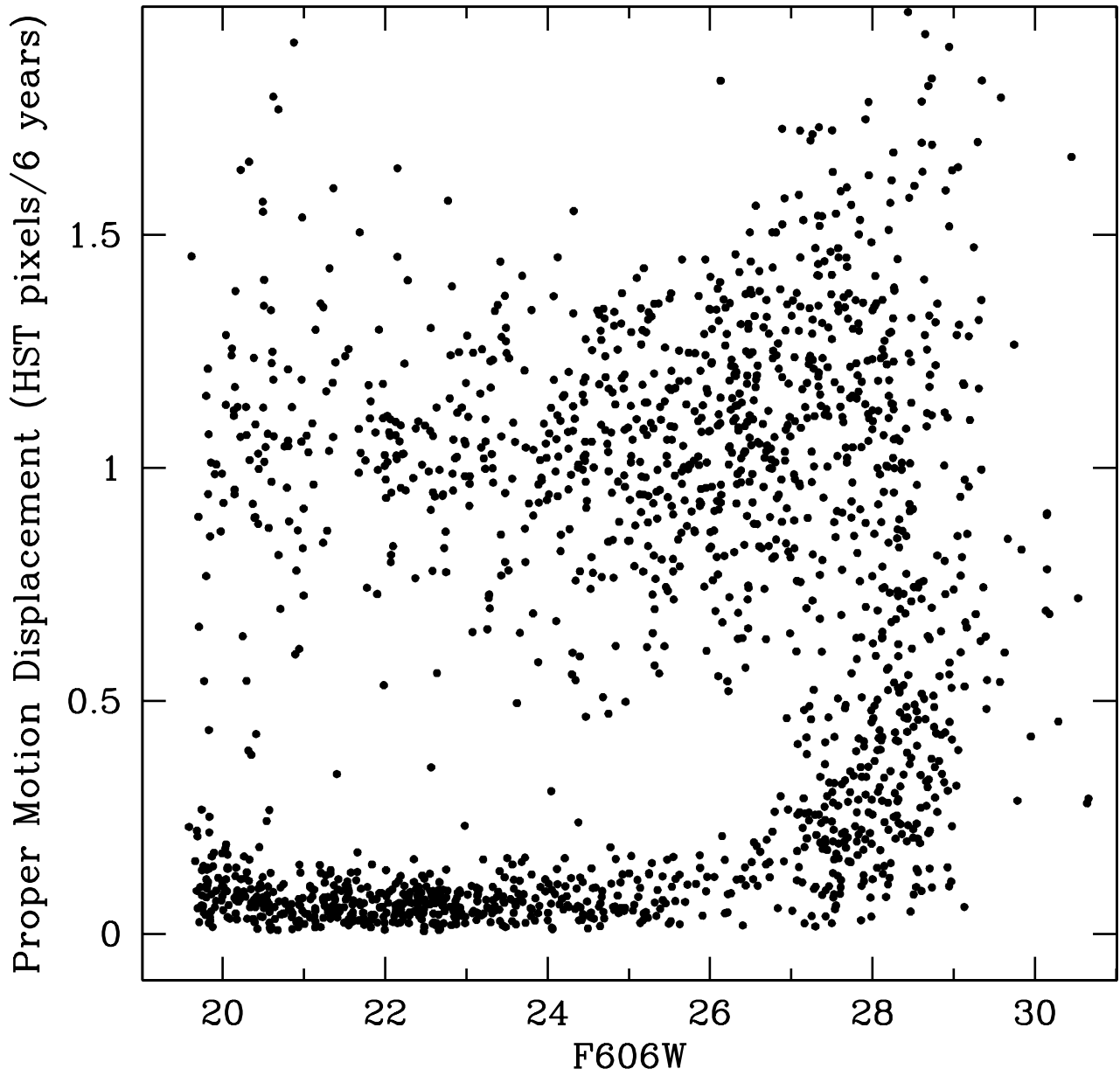
\end{document}